\documentclass[11pt,a4paper]{article}

\usepackage{a4wide}
\usepackage{amsmath,amssymb}
\usepackage{epsfig}

\newcommand{\be}{\begin{equation}}
\newcommand{\ee}{\end{equation}}
\newcommand{\ba}{\begin{eqnarray}}
\newcommand{\ea}{\end{eqnarray}}

\def\bi{\bibitem}

\def\g{\gamma}
\def\no{\nonumber}

\newcommand{\Tr}{{\rm Tr \,}}
\newcommand{\Op}{\mathcal{O}}
\newcommand{\fldZ}{\mathcal{Z}}

\newcommand{\fldD}{\mathcal{D}}

\newcommand{\alg}[1]{\mathfrak{#1}}
\newcommand{\superN}{\mathcal{N}}

\newcommand{\gym}{g\inddowns{YM}}
\newcommand{\gABA}{\gamma\indups{ABA}}
\newcommand{\gRAP}{\gamma\indups{wrapping}}
\newcommand{\EulerGamma}{\gamma_{\indups{E}}}
\newcommand{\indups}[1]{^{\mathrm{\scriptscriptstyle #1}}}
\newcommand{\inddowns}[1]{_{\mathrm{\scriptscriptstyle #1}}}

\begin{document}

\thispagestyle{empty} 

\begin{flushright}\footnotesize
AEI-2009-008 \\
NSF-KITP-09-09
\end{flushright}

 \begin{center}
 \vspace{1.2cm}
{\Large \bf Twist-three at five loops, \\
Bethe Ansatz and wrapping}

 \vspace{1cm} {
  M. Beccaria$~^{a,}$\footnote{matteo.beccaria@le.infn.it}, V.  Forini$~^{b,}$\footnote{
  forini@aei.mpg.de}, T. {\L}ukowski$~^{c,}$\footnote{ tomaszlukowski@gmail.com}, 
  S.  Zieme$~^{b,d,}$\footnote{stefan.zieme@aei.mpg.de}
}\\
 \vskip 0.8cm

{\em
$^{a}$  Physics Department, Salento University and INFN,\\
	 73100 Lecce, Italy \\
 \vskip 0.4cm
$^{b}$ Max-Planck-Institut f\"ur Gravitationsphysik,
        Albert-Einstein-Institut, \\
        Am M\"uhlenberg 1, D-14476 Potsdam, Germany \\
 \vskip 0.4cm
$^{c}$ Institute of Physics, 
	Jagellonian University, \\
	ul. Reymonta 4, 
	30-059 Krak{\'o}w, Poland \\
 \vskip 0.4cm
$^{d}$ Kavli Institute for Theoretical Physics, \\
       University of California Santa Barbara, CA 93106 USA}
\end{center}

 \vskip 3cm

\begin{abstract}
We present a formula for the five-loop anomalous dimension of $\superN=4$  SYM twist-three operators 
in the $\mathfrak{sl}(2)$ sector. We obtain its asymptotic part from the Bethe Ansatz and finite volume corrections
from the generalized L\"uscher formalism, considering scattering processes of spin chain magnons with virtual
particles that travel along the cylinder. The complete result respects the expected large spin scaling properties
and passes non-trivial tests including reciprocity constraints. We analyze the pole structure  and find agreement 
with a conjectured resummation formula. In analogy with the twist-two anomalous dimension at four-loops wrapping 
effects are of order $(\log^2 M/M^2)$ for large values of the spin.
\end{abstract}
%
\renewcommand{\theequation}{1.\arabic{equation}}
\setcounter{equation}{0}

\setcounter{equation}{0}
\setcounter{footnote}{0}
\setcounter{section}{0}

\newpage
\section{Introduction and discussion}
\label{Sec: Intro}

Recent developments in computing finite size effects on the asymptotic spectrum 
of $\superN=4$ SYM twist-two operators are very promising in order to ultimately 
find the complete spectral equations of the dilatation generator.
 
The revealing of integrable structures on both sides of the AdS/CFT correspondence 
\cite{Integrable Structures} gradually led to powerful tools for computing anomalous 
dimension of gauge invariant operators by means of the Bethe Ansatz \cite{Beisert:2005fw}.
The factorized two-body S-Matrix \cite{factorized} that governs scattering processes 
of the spin chain particles and excitations of the $AdS_5 \times S^5$ string worldsheet is determined by the $\alg{psu}(2,2|4)$ symmetry of $\superN=4$ up to a phase factor \cite{Beisert:2005tm}. In order to also determine 
this algebraic ambiguity a crossing-like equation for the dressing phase has been 
derived \cite{Janik:2006dc}. It allows for multiple solutions \cite{Beisert:2006ib}, 
one of which gets singled out \cite{bes} by reconciliation with an explicit diagrammatic 
calculation of the four-loop anomalous dimensions of twist-two
operators in the large spin limit \cite{Bern:2006ew}. It still remains an open problem 
to explicitly show the crossing invariance of the dressed asymptotic Bethe equations.
For this purpose the representations of the dressing factor in \cite{Rej:2007vm} might 
prove useful. 

It was shown that these equations are asymptotic in nature, and need to be corrected 
by wrapping effects \cite{Ambjorn:2005wa}. An explicit calculation of the anomalous
dimension of twist-two operators from the asymptotic Bethe Ansatz at four-loops unequivocally 
showed that the pole prescript by BFKL physics can not be fulfilled \cite{KLRSV}. The Bethe 
Ansatz therefore does not produce the correct result at and beyond wrapping order.

However, for exactly these operators complete results have been obtained for the first time. 
For the simplest representative of twist-two operators, the Konishi-field, a field theoretical 
calculation starting from the asymptotic dilatation generator \cite{Fiamberti:2007rj}, 
finite-size corrections to the Bethe Ansatz using L\"uscher formulas \cite{Bajnok:2008bm}, 
and finally a full-fledged Feynman calculus \cite{Velizhanin:2008jd} identically determined 
the complete anomalous dimension.

The successful application of the L\"uscher formalism relies on a generalization of the L\"uscher 
formulas \cite{Luscher:1985dn} to both non-relativistic models as well as multi-particle states, which had been 
conjectured in \cite{Janikprev,Bajnok:2008bm}.  With this formalism applied to the 2d worldsheet QFT of the $AdS_5\times S^5$ superstring it has been possible to compute the four-loop anomalous dimension of twist-two operators at general values of the spin 
\cite{Bajnok:2008qj}. The result passed several non-trivial tests from BFKL and reciprocity 
constraints \cite{Bajnok:2008qj,bftwist2}. The leading transcendental part had been confirmed 
in an impressive field theory computation \cite{Veliz}.

For the complete spectral equations of $\superN=4$ SYM, however, thermodynamic Bethe Ansatz 
methods ought to be applied, as has been initiated for string and gauge theory in
\cite{Arutyunov:2007tc,Gromov:2008gj}. A Y-system, which is believed to yield anomalous dimensions of arbitrary
local operators of planar $\superN=4$ SYM has been recently conjectured in~\cite{Gromov:2009tv}.

The aim of this work is to continue the application of the  Bethe Ansatz and the L\"uscher 
formalism to the next operators in reach, namely twist-three operators. The leading wrapping contribution to
the anomalous dimension of twist-three operators will appear at five-loops. 
In order to compute the complete five-loop anomalous dimension of the ground state we start from an ansatz based 
on the maximum transcendentality principle~\cite{Kotikov:2002ab} for both the asymptotic 
and wrapping contributions. The asymptotic part can be determined from the Bethe equations
after the initial ansatz has been upgraded with further constraints from reciprocity. 
To compute the wrapping contribution, we apply the generalized L\"uscher formulas~\cite{Bajnok:2008qj} 
to operators of twist-three.

Our result passes some important consistency tests. Its leading asymptotic behavior for large values 
of the spin reproduces the universal scaling function at five-loop order. The first subleading correction 
coincides with the results of~\cite{Freyhult:2009my,Fioravanti:2009xt}. 
Contributions from finite-size effects start at order $(\log^2 M/M^2)$, as in the case of 
twist-two operators~\cite{bftwist2}. In contrast to the latter, there is no BFKL 
equation for twist-three operators, and therefore no prediction for the pole structure of our result. 
However, we analyze the behavior of the anomalous dimension at the singular value of spin
$M=-2$. Interestingly, the pole structure agrees with the 
conjectured resummation formula  of~\cite{KLRSV}, once contributions from wrapping effects are taken 
into account. Additionally, the wrapping correction obtained from the L\"uscher formula precicesly matches 
with the computation from the $Y$-system\footnote{We thank Pedro Vieira for his support in the numerical cross-check.} \cite{Gromov:2009tv}.
The complete result, including the wrapping contribution, is reciprocity respecting.

The main body of the paper follows to the above outlined procedure. Some basic definitions of harmonic sums
are recalled in Appendix~A. Appendix~B contains the analysis of the asymptotic structure of the anomalous 
dimensions and their corresponding $P$-kernels up to five loops. Speculations related to
a pattern in the asymptotic structure of anomalous dimensions of twist operators are also given in Appendix B.
%

\renewcommand{\theequation}{2.\arabic{equation}}
 \setcounter{equation}{0}
\setcounter{equation}{0}

\section{Asymptotic Bethe equations}
We will start our analysis with the contribution to the final result stemming from the asymptotic 
Bethe Ansatz equations. Twist-three operators are embedded in the $\alg{sl}(2)$ sub-sector of $\superN=4$ SYM. 
They can be represented by an insertion of $M$ covariant derivatives $\fldD$ into the protected half-BPS state
$\Tr \fldZ^3$
\begin{equation}\label{eq:twist3}
\Tr \left( \fldD^{s_1}\fldZ\, \fldD^{s_2} \, \fldZ\,\fldD^{s_3}\fldZ\right) + \ldots\,,
\quad  \mathrm{with} \quad M=s_1+s_2+s_3\,. 
\end{equation}
Their anomalous dimensions can be obtained from a non-compact, length-three $\alg{sl}(2)$ spin chain 
with $M$ excitations underlying a factorized two-body scattering \cite{factorized}. However, the 
interaction range between scattering particles increases with orders of the coupling constant in perturbation theory.
If it exceeds the length of the spin chain and {\it wraps} around it, the 
S-matrix picture \cite{factorized,Beisert:2005tm} loses its meaning, as no asymptotic region can be defined any longer. 
For twist $L$ operators this effect, delayed by superconformal invariance, starts at order $g^{2L+4}$. 
Nevertheless, the Bethe Ansatz does not cease to work but gives an incomplete result, which does not 
incorporate these corrections \cite{KLRSV}. 

It was shown that the Bethe Ansatz result for twist-two operators 
can be completed by considering additional scattering effects with virtual particles \cite{Bajnok:2008qj}. 
It passes non-trivial tests with BFKL \cite{Bajnok:2008qj}, as well as  
reciprocity \cite{bftwist2} constraints and reproduces the correct scaling behavior at large values of the spin 
$M$ \cite{bftwist2,bes}, proposing a certain confidence. We compute these wrapping effects for twist $L=3$ 
in section~\ref{Sec: Wrapping}.

The Bethe Ansatz equations for the operators \eqref{eq:twist3} with our choice of the coupling constant 
$g^2=\frac{\gym^2 N}{16\,\pi^2}$ are given by  
\begin{equation}\label{eq:Bethe twist 3}
\left (\frac{x_{k}^{+}}{x_{k}^{-}}\right)^{3}=\prod_{\substack{j=1\\j\neq k}}^{M} 
\frac{x_{k}^{-}-x_{j}^{+}}{x_{k}^{+}-x_{j}^{-}} 
\frac{1-g^{2}/x_{k}^{+}x_{j}^{-}}{1-g^{2}/x_{k}^{-}x_{j}^{+}}\,\exp(2i\theta(u_{k},u_{j}))\,, 
\quad \prod_{k=1}^M \frac{x_k^+}{x_k^-}=1\,.
\end{equation}
The spectral parameters $x^{\pm}$ are defined in terms of the rapidities $u$ by \cite{BDS}
\begin{equation}
x^{\pm}(u)=x(u\pm i/2)\,, \quad x(u)=\frac{u}{2}\left(1+\sqrt{1-{4g^2}/{u^2}}\right)\,.
\end{equation}
The phase shift $\theta(u_{k},u_{j})$ acquired by two particles with rapidities $u_k,u_j$ passing 
each other is given to five-loop order by  the expansion \cite{bes}  
\begin{equation}
 \theta(u_{k},u_{j})=\left(4 \zeta_3 g^6 - 40 \zeta_5 g^8 \right) 
\left(q_2(u_k)q_3(u_j)-q_3(u_k)q_2(u_j)\right)+ \Op(g^{10})\,, 
\end{equation}
where the $q_r(u)$ correspond to the conserved magnon charges \cite{BDS}
\begin{equation}
 q_r(u) = \frac{i}{r-1} \left( \frac{1}{(x^+(u))^{r-1}}-\frac{1}{(x^{-}(u))^{r-1}} \right)\,.
\end{equation}
From the solution to \eqref{eq:Bethe twist 3} given in terms of the Bethe roots $x^{\pm}_k$ 
one computes the anomalous dimension by 
\begin{eqnarray}
 \gABA(g)=2g^2 \sum_{k=1}^{M} q_2(u_k)\,.
\end{eqnarray}
Due to the length $L=3$ of the operators \eqref{eq:twist3}, wrapping effects are expected 
to contribute at order $g^{10}$, such that the anomalous dimension is written perturbatively as 
 \ba\label{gamma}
\g(M) &=& g^2\,\g_2(M)+g^4\,\g_4 (M)+g^6\,\g_6 (M)+g^8\,\g_8 (M)\\
&& + g^{10}\,(\gABA_{10} (M)+\gRAP_{10}(M))+\ldots\nonumber \,,
\ea
where we tacitly assumed that to order $g^8$ the complete result is identical to $\gABA$ and therefore 
dropped its index.

At one-loop the Bethe roots $u_k$ are given by zeros of the Wilson polynomial \cite{KLRSV}
\begin{equation}
\label{eq:Wilsonpolynomial}
	P_M(u)={}_4 F_3\left(\left. \begin{array}{c}
	-\frac{M}{2}, \ \frac{M}{2}+1,\ \frac{1}{2}+iu ,\ \frac{1}{2}-iu\\
	1,\  1 ,\ 1 \end{array}
	\right| 1\right) \,.
\end{equation}
Closed expressions for the corrections to the Bethe roots to three-loop order have also been obtained 
 in \cite{Kotikov:2008pv} from the Baxter approach \cite{Belitsky:2006wg}. However, it is currently 
unclear if the asymptotic Baxter equation \cite{Belitsky:2006wg} reproduces the same result as 
the Bethe Ansatz at and beyond wrapping order. 

In order to obtain closed expressions for the anomalous dimension we will therefore solve 
\eqref{eq:Bethe twist 3} perturbatively for fixed values of the spin $M$ and match the coefficients 
in an appropriate ansatz which assumes the maximum transcendentality principle~\cite{Kotikov:2002ab}. 
Up to four loops, these expressions have been derived in~\cite{KLRSV} and~\cite{Beccaria:2007cn}. 
They are given by 
\begin{eqnarray}\label{g2}
\g_2&=&8\,S_1\\\label{g4}
\g_4&=&-16\, S_1\, S_2-8\, S_3\\
\g_6&=& 32\, S_1\, S_2^2+48 \, S_3\, S_2+16\, S_1\, S_4+40\, S_5-32 \,S_{2,3}
      +64\, S_1\, S_{3,1}+32\, S_{4,1}-64 \, S_{3,1,1}\\\label{g8}
\g_8&=&8 \,S_{7} + 112\, S_{1, 6} + 240 \,S_{2, 5} - 80\, S_{3, 4} - 464\, S_{4, 3} - 
 336 \,S_{5, 2} - 80 \,S_{6, 1} - 640 \,S_{1, 1, 5} \\\no
 &&- 512\, S_{1, 2, 4}+ 
 384\, S_{1, 3, 3} + 512\, S_{1, 4, 2} - 512\, S_{2, 1, 4} + 320 \,S_{2, 2, 3} + 
 640\, S_{2, 3, 2} + 64 \, S_{2, 4, 1} \\\no
 &&+ 384 \, S_{3, 1, 3}+ 704 \,S_{3, 2, 2}+ 384\, S_{3, 3, 1} + 576 \, S_{4, 1, 2} + 576\, S_{4, 2, 1} + 384 \, S_{5, 1, 1} + 
 1280 \, S_{1, 1, 1, 4} \\\no
 &&- 256\, S_{1, 1, 3, 2} + 512 \, S_{1, 1, 4, 1}- 384 \, S_{1, 2, 2, 2} + 256 \, S_{1, 2, 3, 1} - 384 \, S_{1, 3, 1, 2} - 
 384 \, S_{1, 3, 2, 1}\\\no
 && - 384 \, S_{1, 4, 1, 1} - 384 \, S_{2, 1, 2, 2} + 
 256 \, S_{2, 1, 3, 1}- 384 \, S_{2, 2, 1, 2} - 384\, S_{2, 2, 2, 1} - 
 384 \, S_{2, 3, 1, 1}\\\no
 && - 384 \, S_{3, 1, 1, 2} - 384 \, S_{3, 1, 2, 1} - 
 384 \, S_{3, 2, 1, 1} - 384 \, S_{4, 1, 1, 1} - 1024 \, S_{1, 1, 1, 3, 1}- 128 \,S_1\,S_3\,\zeta_3 \,.
\end{eqnarray}
All sums are evaluated at argument $M/2$ and only positive indices appear.
At four loop-order, the dressing phase of the Bethe equations  
starts to contribute to (\ref{g8}) with a term proportional to $\zeta_3$.

To determine the five-loop result in the same fashion implies a tremendous computational 
effort in view of the necessary precision~\footnote{See~\cite{Beccaria:2007gu} for the five-loop anomalous dimension of a different class of operators (the field strength operators Tr$\,{\mathcal{F}^L}$),  determined as a closed function of their length $L$.}. 
We have obtained rational values for the anomalous 
dimension up to $M\approx200$, which is however too far from the requirement to fit the coefficients 
of the corresponding ansatz vector of constant degree of transcendentality. In order to find 
a closed form for $\gABA_{10}$ in terms of nested harmonic sums further constraints to reduce the 
number of coefficients are needed.

\renewcommand{\theequation}{3.\arabic{equation}}
 \setcounter{equation}{0}
 
\section{Parity invariance of $\gamma(M)$}

The multi-loop anomalous dimension $\gamma(M)$ is conjectured to obey a powerful constraint known as  generalized Gribov-Lipatov \emph{reciprocity}. This constraint, arising in the QCD context, has been presented in~\cite{dok1,dok2} as a special (space-time symmetric) reformulation of the parton distribution functions evolution equation, and approached in~\cite{bk} from the point of view
of the large $M$ expansion. In particular, in~\cite{bk} such analysis has been generalised to anomalous dimensions of operators of arbitrary twist $L$, and reciprocity has been dubbed \emph{parity invariance}
in the sense clarified below. Reciprocity  has been checked in various multi-loop calculations of weakly coupled  ${\cal N}=4$ gauge theory~\cite{bf,Forini:2008ky,bdm,Beccaria:2007pb,Beccaria:2007cn,bftwist2}. 

The reciprocity or parity invariance condition 
is easily expressed in terms of the $P$-function (kernel), depending on the Lorentz spin $M$,
which is  in one-to-one correspondence, at least perturbatively, with the anomalous dimension $\gamma(M)$ as follows from~\cite{dok1,bk,dok2}
\be
\label{nonlinear}
\gamma(M) = P\left(M+\textstyle{\frac{1}{2}\gamma(M)}\right).
\ee
The parity invariance condition is a constraint that arises in the large $M$ expansion of $P(M)$, which is expected to take the 
following form
\be
\label{RRJ}
\qquad P(M) = \sum_{\ell\ge 0} \frac{a_\ell(\log\,J^2)}{J^{2\,\ell}}, \qquad J^2=\frac{M}{2}\,\left(\frac{M}{2}+1\right),
\ee  
where the $a_\ell$ are {\em coupling-dependent} polynomials. Eq.~(\ref{RRJ}) implies an infinite 
set of constraints on the coefficients of the large $M$ expansion of $P(M)$ organized in a standard $1/M$ 
power series. The name {\em parity-invariance} is related to the absence of terms of the form 
$1/J^{2n+1}$, odd under $J\to -J$.

In the following we will use the constraint Eq.~(\ref{RRJ}) as a guiding principle in order to 
obtain the five-loop expression $\gABA_{10}$. To this aim, we need to express it as a more practical 
test such that it can be applied to any proposed combination of harmonic sums. This task can be performed with a 
basic result of~\cite{bftwist2}, which we recall in following.
%
\subsection{Harmonic combinations with definite parity}

The notation for complementary harmonic sums $\underline{S}_\mathbf{a}$ is recalled in Appendix A. 
Let us introduce the map $\omega_a$, $a\in \mathbb{N}$, which acts linearly on linear 
combinations of harmonic sums 
\be
\omega_a(S_{b, \mathbf{c}}) = S_{a, b, \mathbf{c}}-\frac{1}{2}\,S_{a+b, \mathbf{c}}.
\ee
We also introduce a complementary map $\underline\omega_a$, acting in a similar way on complementary sums
\be
\underline\omega_a(\underline {S_{b, \mathbf{c}}}) = \underline {S_{a, b, \mathbf{c}}}-\frac{1}{2}\,\underline{ S_{a+b, \mathbf{c}}}.
\ee
Finally, let us introduce the combinations $\Omega$ 
\ba
\Omega_a &=& S_a,  \\
\Omega_{a, \mathbf{b}} &=& \omega_a\,(\Omega_\mathbf{b}),
\ea
and the analogous complementary combinations $\underline\Omega_\mathbf{a}$. It is  of course 
possible to change the basis from $S_\mathbf{a}$ to $\underline\Omega_\mathbf{a}$. 
For example, up to a degree of transcendentality three, we have 
\ba
\underline\Omega_1 &=& S_1\,, \quad \underline\Omega_2 = S_2\,, \quad \underline{\Omega_{1,1}} = S_{1,1}-\frac{S_2}{2}\,, 
\quad \underline\Omega_3= S_3\,,\quad \underline{\Omega_{2,1}} = S_{2,1}-\frac{S_3}{2}\,, \no\\
\underline{\Omega_{1,2}} &=& -\frac{1}{6} \pi ^2 S_1-\frac{S_3}{2}+S_{1,2}\,, \quad \underline{\Omega_{1,1,1}} = \frac{S_3}{4}-\frac{S_{1,2}}{2}-\frac{S_{2,1}}{2}+S_{1,1,1}\,,
\ea
which can be inverted in order to obtain 
\ba
S_2     &=& \underline\Omega_2\,, \quad S_{1,1} = \frac{\underline\Omega_2}{2}+\underline{\Omega_{1,1}}\,, \quad 
S_3       = \underline\Omega_3\,, \quad S_{2,1}   = \frac{\underline\Omega_3}{2}+\underline{\Omega_{2,1}}, \no\\
S_{1,2}   &=& \frac{\pi ^2
   \underline\Omega_1}{6}+\frac{\underline\Omega_3}{2}+\underline{\Omega_{1,2}}\,, \quad
S_{1,1,1} = \frac{\pi ^2\, \underline{\Omega_1}}{12}+\frac{\underline\Omega_3}{4}+\frac{\underline{\Omega_{1,2}}}{2}+\frac{\underline{\Omega_{2,1}}}{2}+\underline{\Omega_{1,1,1}}.
\ea

The crucial result is then given by the following theorem~\cite{bftwist2}.

\noindent{\bf Theorem}: 

\noindent {\em (a) The combination $\underline\Omega_{a_1, \dots, a_d}$ with positive $\{a_i\}$
is parity-even iff 
\be
(-1)^{a_1 + \cdots + a_d} = (-1)^d.
\ee

\noindent (b) If this condition is not satisfied, the expansion of $\underline\Omega_{a_1, \dots, a_d}$ 
is parity-odd,  with the (trivial) exception of the  leading constant term.

\noindent
(c) The combination $\Omega_{a_1, \dots, a_d}$ with positive odd $\{a_i\}$ is parity-even.}

\bigskip
From this theorem we deduce the following 

\noindent{\bf Theorem (parity-invariance test): }{\em a specific linear combination of 
harmonic sums is parity invariant iff it does not contain parity-odd terms when transformed 
from the $S_\mathbf{a}$ basis to the $\underline\Omega_\mathbf{a}$ basis.}

To see how this test can be used let us consider an illustrative example, the two-loop 
anomalous dimension. One starts with the following ansatz of transcendentality three  
\be
\g_4=a_1 S_3+a_2 \,S_{1,2}+a_3\,S_{2,1}+a_4\,S_{1,1,1}\,,
\ee
with all sums evaluated at $M/2$. The corresponding $P_{4}$-kernel, derived by 
inverting formula (\ref{nonlinear}) and replacing the perturbative expansion (\ref{gamma}), 
reads in a canonical basis
\be\label{P_4}
P_4=\gamma_4-\frac{1}{4}\g_2\g_2'\equiv(a_1-16) S_3+(a_2+16) S_{1,2}+(a_3+16) S_{2,1}-16\,\zeta_2\, S_1+a_4S_{1,1,1}\,,
\ee
where we used the one-loop result (\ref{g2}). Writing (\ref{P_4}) in terms of the 
$\underline\Omega$ basis one finds
\be\label{P4bis}
P_4=c_1\,\underline{\Omega_{1}}+c_3\,\underline{\Omega_{3}}+c_{1,2}\,\underline{\Omega_{1,2}}+c_{2,1}\,\underline{\Omega_{2,1}}+c_{1,1,1}\,\underline{ \Omega_{1,1,1}}
+{\rm const}\,,
\ee
where the $c_i$ are linear combinations of the coefficients $a_i$. The combinations 
$\underline{\Omega_{1}},\,\,\underline{\Omega_{3}},\,\,\underline{\Omega_{1,1,1}}$ 
are all reciprocity respecting, according to the above theorem. Imposing reciprocity 
on $P_4$ implies the vanishing of the coefficients of those $\underline\Omega$ 
with wrong parity, namely 
\be
c_{1,2}=a_2+16+\frac{a_4}{2}=0,~~~~~~~~~~~~~~~c_{2,1}=a_3+16+\frac{a_4}{2}=0\,.
\ee
This leads to the conditions $a_3=a_2$ and $a_4=-2(16+a_2)$, that are indeed satisfied 
by the expression in \eqref{g4}. Thus, reciprocity has determined $2$ of the $4$ unknown 
coefficients in the initial ansatz for the anomalous dimension~\footnote{The coefficient $a_4$ 
has only been kept to show the exact number of constraints coming from reciprocity. It could have 
been set to zero from the beginning because at large $M$ the term $S_{1,1,1}\sim \log^3M$ is not 
compatible with the universal leading logarithmic behavior (cusp anomaly).}.
%
\renewcommand{\theequation}{4.\arabic{equation}}
 \setcounter{equation}{0}

\section{Determination of $\gABA_{10}$}

The strategy to derive the asymptotic part of the anomalous dimension at  $n=5$ loops incorporates a 
combined use of the maximum transcendentality principle, reciprocity and Bethe equations.

The starting point  is to write $\gABA_{10}$ as a linear combination of harmonic sums of 
transcendentality $\tau=2n-1=9$. For a given $\tau$, basic combinatorics leads to the fact that 
there are $2^{\tau-1}$ linearly independent harmonic sums with positive indices.
This means that there are in principle $256$ terms which potentially contribute to the anomalous dimension. 

From the numerical solution of the asymptotic Bethe equations it is possible to obtain a long 
list of rational values for $\gABA_{10}(M)$ for fixed values of $M$. The list-length is smaller 
than 256 due to rather hard computational limitations. However, these limitations can be overcome 
by means of parity-invariance. 

To constrain the $256$ unknown coefficients via reciprocity one has to impose parity in the sense 
of Eq.~(\ref{RRJ}) on the five-loop contribution $P_{10}$ to the kernel $P$ defined in Eq.~(\ref{nonlinear}). 
This contribution can be derived from the anomalous dimension by simply inverting Eq.~(\ref{nonlinear}) 
and taking into account the perturbative expansion Eq.~(\ref{gamma}). Finally, we apply the previous 
parity-invariance test and obtain a large set of linear constraints on the unknown coefficients.
The total number of constraints from Bethe equations and parity-invariance is now larger than $256$ 
and we find an over-determined set of linear equations, which is solvable. The final 
result is given in Table \ref{fiveloops}, in which terms multiplied by $\zeta_3$ and $\zeta_5$ are
directly induced from the dressing factor. As is the case for lower-loop orders,
only positive indices appear in the participating harmonic sums.
The result that we have obtained by the above stated methods can be checked as follows:

\begin{enumerate}
\item {\em Scaling function (cusp anomaly)}. A consistency check of the formula presented 
in Table~\ref{fiveloops} is given by its leading asymptotic behavior, namely
\be\label{g10asy}
\gABA_{10}(M)\sim 32\Big(\frac{887}{14175}\pi^8+\frac{4}{3}\pi^2\,\zeta_3^2+40\,\zeta_3\,\zeta_5\Big) \log M, \quad  \mathrm{for} \quad M \to \infty \,.
\ee
It coincides with the five-loop contribution in the weak coupling perturbative expansion of 
the integral equation obtained in \cite{bes}, which is believed to describe 
the universal scaling function.
Put differently,
we confirm its universality~\cite{Belitsky:2003ys,Eden:2006rx} for twist-three at five loops.

\section*{Asymptotic five-loop anomalous dimension of twist-three operators}
\begin{table}[hp!]
\begin{eqnarray}\no
&&{\bf{\gamma}}\indups{{\bf{ABA}}}_{\bf 10}
=136 S_ {9} + 368  S_ {1, 8} + 2832 S_ {2, 7} + 4272   
S_ {3, 6} + 848   S_ {4, 5} - 3024 S_ {5, 4} - 2736  
S_ {6, 3} - 1168 S_ {7, 2} \\\no
 &&- 496  S_ {8, 1} - 5376  
S_ {1, 1, 7} - 12352  S_ {1, 2, 6} - 8832  S_ {1, 3, 5} + 1600  S_ {1, 4, 4} + 3968  S_ {1, 5, 3} - 64   S_ {1, 6, 2} \\\no
 &&- 1344  
S_ {1, 7, 1} - 12352   S_ {2, 1, 6} - 13760   
S_ {2, 2, 5} - 2112   S_ {2, 3, 4} + 4288 S_ {2, 4, 3}- 960  
S_ {2, 5, 2} - 5440  S_ {2, 6, 1}\\\no
 && - 9088  S_ {3, 1, 5} - 2432  
S_ {3, 2, 4} + 5120   S_ {3, 3, 3} + 2688   S_ {3, 4, 2} - 4160  
S_ {3, 5, 1} + 1280   S_ {4, 1, 4} + 5824   S_ {4, 2, 3}  \\\no
 &&+6400  
S_ {4, 3, 2} + 2112  S_ {4, 4, 1} + 5120  S_ {5, 1, 3} + 6208  
S_ {5, 2, 2} + 5312  S_ {5, 3, 1} + 3904   S_ {6, 1, 2} + 3904  
S_ {6, 2, 1} \\\no
 &&+ 1728  S_ {7, 1, 1} + 21504 
S_ {1, 1, 1, 6} + 22784  S_ {1, 1, 2, 5} + 5632  
S_ {1, 1, 3, 4} - 1280   S_ {1, 1, 4, 3} + 6912  
S_ {1, 1, 5, 2} \\\no
 &&+ 11520   S_ {1, 1, 6, 1} + 22784  
S_ {1, 2, 1, 5} + 9088 S_ {1, 2, 2, 4} - 1024  
S_ {1, 2, 3, 3} + 6784  S_ {1, 2, 4, 2} + 17152  
S_ {1, 2, 5, 1} \\\no
 &&+ 5504 S_ {1, 3, 1, 4} - 3456   
S_ {1, 3, 2, 3} - 1536   S_ {1, 3, 3, 2} + 7680  
S_ {1, 3, 4, 1} - 4480  S_ {1, 4, 1, 3} - 6272 
S_ {1, 4, 2, 2}\\\no
 &&- 3584 S_ {1, 4, 3, 1} - 3840 
S_ {1, 5, 1, 2} - 3840  S_ {1, 5, 2, 1} + 768 
S_ {1, 6, 1, 1} + 22784   S_ {2, 1, 1, 5} + 9088  
S_ {2, 1, 2, 4}\\\no
 &&- 1024  S_ {2, 1, 3, 3} + 6784  
S_ {2, 1, 4, 2} + 17152 S_ {2, 1, 5, 1} + 9088  S_ {2, 2, 1, 4} - 2688 \, S_ {2, 2, 2, 3} + 640 \, 
S_ {2, 2, 3, 2} \\\no
 &&+ 13440   S_ {2, 2, 4, 1} - 3456  
S_ {2, 3, 1, 3} - 7040 S_ {2, 3, 2, 2} - 768  
S_ {2, 3, 3, 1} - 4480  S_ {2, 4, 1, 2} - 4480 S_ {2, 4, 2, 1} \\\no
 &&+ 2816  S_ {2, 5, 1, 1} + 6272  
S_ {3, 1, 1, 4} - 2944  S_ {3, 1, 2, 3} - 1536 
S_ {3, 1, 3, 2} + 7936  S_ {3, 1, 4, 1}- 2944   
S_ {3, 2, 1, 3} \\\no
 && - 7296  S_ {3, 2, 2, 2} - 768  
S_ {3, 2, 3, 1} - 6656  S_ {3, 3, 1, 2} - 6656  
S_ {3, 3, 2, 1} - 1024  S_ {3, 4, 1, 1} - 3968  
S_ {4, 1, 1, 3} \\\no
 &&- 6528   S_ {4, 1, 2, 2} - 3584 
S_ {4, 1, 3, 1} - 6528   S_ {4, 2, 1, 2} - 6528 
S_ {4, 2, 2, 1} - 4864   S_ {4, 3, 1, 1} - 5376  
S_ {5, 1, 1, 2} \\\no
 &&- 5376  S_ {5, 1, 2, 1} - 5376 
S_ {5, 2, 1, 1} - 4608   S_ {6, 1, 1, 1} - 32768  
S_ {1, 1, 1, 1, 5} - 10240   S_ {1, 1, 1, 2, 4} - 3072  
S_ {1, 1, 1, 3, 3} \\\no
 &&- 17920   S_ {1, 1, 1, 4, 2} - 30720  S_ {1, 1, 1, 5, 1} - 10240   S_ {1, 1, 2, 1, 4} - 8704 
S_ {1, 1, 2, 3, 2}- 24064 S_ {1, 1, 2, 4, 1} \\\no
 && + 1024  
S_ {1, 1, 3, 1, 3} + 2560   S_ {1, 1, 3, 2, 2} - 4096  
S_ {1, 1, 3, 3, 1} - 512 S_ {1, 1, 4, 1, 2} - 512   
S_ {1, 1, 4, 2, 1}- 10240   S_ {1, 1, 5, 1, 1} \\\no
 && - 10240 
S_ {1, 2, 1, 1, 4} - 8704   S_ {1, 2, 1, 3, 2} - 24064  
S_ {1, 2, 1, 4, 1}+ 3072  S_ {1, 2, 2, 2, 2}  - 6656   
S_ {1, 2, 2, 3, 1} \\\no
 &&+ 512  S_ {1, 2, 3, 1, 2} + 512 
S_ {1, 2, 3, 2, 1} - 10752  S_ {1, 2, 4, 1, 1}+ 1024 
S_ {1, 3, 1, 1, 3} + 3072  S_ {1, 3, 1, 2, 2}  - 3584 
S_ {1, 3, 1, 3, 1}  \\\no
 && + 3072 S_ {1, 3, 2, 1, 2} + 3072 
S_ {1, 3, 2, 2, 1} - 2560  S_ {1, 3, 3, 1, 1}+ 3072  
S_ {1, 4, 1, 1, 2}   + 3072   S_ {1, 4, 1, 2, 1} + 3072   
S_ {1, 4, 2, 1, 1}\\\no
 && + 3072 S_ {1, 5, 1, 1, 1} - 10240 
S_ {2, 1, 1, 1, 4} - 8704  S_ {2, 1, 1, 3, 2}- 24064  
S_ {2, 1, 1, 4, 1} + 3072   S_ {2, 1, 2, 2, 2} \\\no
 &&- 6656 
S_ {2, 1, 2, 3, 1}  + 512  S_ {2, 1, 3, 1, 2} + 512  
S_ {2, 1, 3, 2, 1}  - 10752 S_ {2, 1, 4, 1, 1} + 3072  
S_ {2, 2, 1, 2, 2} - 6656  S_ {2, 2, 1, 3, 1}  \\\no
 &&+ 3072 
S_ {2, 2, 2, 1, 2} + 3072   S_ {2, 2, 2, 2, 1}  - 5632  
S_ {2, 2, 3, 1, 1} + 3072  S_ {2, 3, 1, 1, 2} + 3072  
S_ {2, 3, 1, 2, 1} + 3072  S_ {2, 3, 2, 1, 1} \\\no
 &&+ 3072  
S_ {2, 4, 1, 1, 1}  + 3072 S_ {3, 1, 1, 2, 2} - 4096 
S_ {3, 1, 1, 3, 1} + 3072   S_ {3, 1, 2, 1, 2} + 3072  
S_ {3, 1, 2, 2, 1} - 2560   S_ {3, 1, 3, 1, 1}  \\\no
 &&+ 3072 
S_ {3, 2, 1, 1, 2} + 3072  S_ {3, 2, 1, 2, 1} + 3072  
S_ {3, 2, 2, 1, 1} + 4608   S_ {3, 3, 1, 1, 1}  + 3072 
S_ {4, 1, 1, 1, 2}  + 3072 S_ {4, 1, 1, 2, 1} \\\no
 &&+ 3072 
S_ {4, 1, 2, 1, 1} + 3072  S_ {4, 2, 1, 1, 1} + 3072 
S_ {5, 1, 1, 1, 1} + 16384   S_ {1, 1, 1, 1, 3, 2}+ 32768  
S_ {1, 1, 1, 1, 4, 1} \\\no
 &&+ 8192   S_ {1, 1, 1, 2, 3, 1}   + 4096 
S_ {1, 1, 1, 3, 1, 2} + 4096   S_ {1, 1, 1, 3, 2, 1}  + 20480 
S_ {1, 1, 1, 4, 1, 1} + 8192 \, S_ {1, 1, 2, 1, 3, 1} \\\no
 &&+ 12288 
S_ {1, 1, 2, 3, 1, 1}  + 8192  S_ {1, 2, 1, 1, 3, 1}  + 12288  
S_ {1, 2, 1, 3, 1, 1} + 8192   S_ {2, 1, 1, 1, 3, 1} + 12288  
S_ {2, 1, 1, 3, 1, 1} \\\no
 &&- 16384  
S_ {1, 1, 1, 1, 3, 1, 
   1}  + {\bf \zeta_{\bf 3}}\,(896   S_ {6} - 2304  S_ {1, 5} - 1792  
    S_ {2, 4} - 768 S_ {3, 3} - 1792   S_ {4, 2} - 2304   
    S_ {5, 1} \\\no
 &&+ 2560   S_ {1, 1, 4} + 512 S_ {1, 2, 3}  + 1536  
    S_ {1, 3, 2} + 3584   S_ {1, 4, 1} + 512 
    S_ {2, 1, 3} + 1536  S_ {2, 3, 1} + 512  
    S_ {3, 1, 2} \\\no
 && + 512  S_ {3, 2, 1}   + 2560 
    S_ {4, 1, 1} - 2048  S_ {1, 1, 3, 1}- 2048 
    S_ {1, 3, 1, 1}) +\, 1280 \, {\bf{\zeta}}_{\bf{5}}\,(S_ {1,3} +S_{3,1}-S_4 )
\nonumber
\end{eqnarray}
\caption{The result for the five-loop asymptotic dimension
$\gABA _{10} \big(\textstyle{\frac{M}{2}}\big)$, written in the canonical basis. }
\label{fiveloops}
\end{table}

\item{\em Virtual scaling function.} 
A further confirmation can be found from the evaluation of the first finite-order correction 
to the asymptotic behavior (\ref{g10asy}). In general this quantity is twist-dependend
and thus non-universal~\cite{Freyhult:2007pz}. However, it has been shown that this
dependence is only linear and an all-loop integral equation can be written~\cite{Freyhult:2009my} (see also~\cite{Bombardelli:2008ah,Fioravanti:2009xt}).
The large $M$ expansion of Table~\ref{fiveloops} leads to the $\Op(1/M^0)$ value
\be 
B_3^{(5)}=\frac{2048}{945} \pi ^6 \zeta_3+64 \zeta_3^3+\frac{8}{45} \pi ^4 \zeta_5-\frac{440}{3} \pi ^2 \zeta_7-7448 \zeta_9 \,.
\ee
It coincides with the expression written explicitly in~\cite{Fioravanti:2009xt}.
Further interesting observations on the other subleading terms in the asymptotic 
expansion of $\gABA_{10}$  will be discussed in Appendix~B.

\item {\em BFKL-like poles}. An indirect indication of the correctness 
of the result emerges by looking at its analytical continuation to complex values of the spin.
In particular, the structure of the expansion around $M=-2$ will be presented in Section~\ref{Sec:AnCont}. 

\item {\em Dressing self-consistency}. The dressing induced terms in $\gABA_{10}$ 
are separately parity invariant. The dressing factor starts to contribute at four loops.
It was observed that at this order terms of the anomalous dimensions of twist-two  
and -three operators coming from the dressing factor are reciprocity respecting 
\emph{separately}~\cite{bdm,bf, bftwist2}.
The analysis of $P_{10}$ in the case of $\gABA_{10}$ confirms this feature. 
The five-loop term proportional to $\zeta_3$  is reciprocity respecting if combined with the corresponding
four-loop $\zeta_3$-term. The  $\zeta_5$-term at five loops is reciprocity
respecting separately (see formula~\ref{P5RR}). This seems to indicate a perturbative pattern 
for the reciprocity of terms that are dressing-induced. Terms proportional to transcendental 
sums $\zeta_i$, which newly appear at a given loop order should automatically be reciprocity 
respecting. Terms proportional to transcendental sums that are also present at lower-loop orders 
are invariant under (\ref{RRJ}) when combined altogether.

\item {\em Additional structural properties}. All coefficients of the harmonic
sums are integers, likewise to the lower loop orders. Also, $P_{10}$ turns out to be a combination 
of allowed parity-even combinations of type  $\Omega$, a condition being stronger 
than the general parity invariance. 
\end{enumerate}
%
\renewcommand{\theequation}{5.\arabic{equation}}
 \setcounter{equation}{0}

\section{The wrapping contribution}
\label{Sec: Wrapping}

In this section we evaluate the leading wrapping correction to the asymptotic 
anomalous dimension of twist-three operators.

The L\"uscher type formula for multi-particle states was conjectured in~\cite{Bajnok:2008bm} 
and successfully applied to the Konishi-operator in~\cite{Bajnok:2008bm} as well as  
twist-two operators of general spin in~\cite{Bajnok:2008qj}. It consists of two parts. One describes the 
modification of the particle quantization condition due to the finite volume (which will 
not contribute at leading order), while the second comes from the propagation of virtual 
particles around the cylinder and is given by
	\begin{equation}
	\Delta E(L)=-\sum_{Q=1}^\infty \int_{-\infty}^{\infty}\frac{dq}{2\pi} \mbox{STr}_{a_{1}} 
	\left[S_{a_{1}a}^{a_{2}a}(q,p_{1})S_{a_{2}a}^{a_{3}a}(q,p_{2})\dots S_{a_{M}a}^{a_{1}a}(q,p_{M})\right]
	e^{-\tilde{\epsilon}_{a_{1}}(q)L} \,.\label{eq:Luscher}
	\end{equation}
This formula applies to a $M$-particle state of identical particles of type $a$, whose
consecutive self-scatterings preserve the state and determine their momenta $p_{i}$ via the 
ABA equations. The matrix $S_{ba}^{ca}(q,p)$ describes how a virtual particle of type $b$ with 
momentum $q$ scatters on a real particle of type $a$ with momentum $p$. The exponential factor 
can be interpreted as the propagator of the virtual particle.

For twist-three operators the momenta of the particles are determined by the ABA equations
in terms of the rapidities $u$ by
	$$
	u(p)=\frac{1}{2}\cot\frac{p}{2}\sqrt{1+16g^{2}\sin^{2}\frac{p}{2}}\,\,.
	$$
At one-loop the rapidities are given by roots of the Baxter-$Q$ function $P_M(u)$ in 
\eqref{eq:Wilsonpolynomial}. As this is an even polynomial of order $M$, we can 
repeat the derivation of~\cite{Bajnok:2008qj}, which leads to a result similar to the 
twist-two case. However, we have to take into account two differences. 
The first one is that the length is equal to $L=3$, which renders 
the exponential part to be of the form
	$$
	e^{-\tilde{\epsilon}_{Q}(q)L}=\frac{4^{L}g^{2L}}{(q^{2}+Q^{2})^{L}}\stackrel{L=3}{=}\frac{64g^6}{(q^{2}+Q^{2})^{3}}\,.
	$$
Additionally, the one-loop energy of twist-three operators differs from the 
the twist-two one. It is given by  
	$$
	\sum_{k=1}^{M}\frac{16}{1+4u_{k}^{2}}=8S_{1}\big(\tfrac{M}{2}\big)\,.
	$$
In the end, we can write the wrapping correction in a very elegant way as
	\begin{equation}\label{eq:wrappingintegral}
	\Delta \gamma=-64g^{10}\, S_{1}\Big( \frac{M}{2}\Big) ^{2}\,\sum_{Q=1}^{\infty}\int_{-\infty}^{\infty}\frac{dq}{2\pi}
	\frac{T_M(q,Q)^{2}}{R_M(q,Q)}\frac{64}{(q^{2}+Q^{2})^{3}}~,
	\end{equation}
where $R_M$ and $T_M$ are functions given by the same expressions that are valid in the case of 
twist-two operators
	\begin{eqnarray*}
	R_M(q,Q)&=&P_M\big(\tfrac{1}{2}(q-i(Q-1))\big)P_M\big(\tfrac{1}{2}(q+i(Q-1))\big) \no\\
	&&\times P_M\big(\tfrac{1}{2}(q+i(Q+1))\big) P_M\big(\tfrac{1}{2}(q-i(Q+1))\big)\,,\\
	T_M(q,Q)&=&\sum_{j=0}^{Q-1}\Big[\frac{1}{2j-iq-Q}-\frac{1}{2(j+1)-iq-Q}\Big]
	P_M\left(\tfrac{1}{2}(q-i(Q-1))+ij\right)\,.
	\end{eqnarray*}

In order to obtain the wrapping contribution we calculated (\ref{eq:wrappingintegral}) for all 
even values of $M$ up to $M=40$. Assuming the maximal transcendentality principle, we expect the 
wrapping correction to have the following structure
	\begin{equation}\label{eq:wrappingtranscendental}
	\gRAP(M)=S_{1}\big(\tfrac{M}{2}\big)^2(C_{0}(M)\zeta_7+C_{2}(M)\zeta_5+C_{4}(M)\zeta_3+C_{7}(M))\,,
	\end{equation}
where the coefficients $C_{n}(M)$ have a degree of transcendentality $n$. We used the fact 
that $S_1^2$ is factored out in the L\"uscher formula (\ref{eq:wrappingintegral}). Likewise to 
the case of the asymptotic Bethe Ansatz we are looking for coefficients that are linear combination of 
harmonic sums with positive indices. For a given degree of transcendentality $n$, there are $2^{n-1}$ 
independent sums. Thus, in order to obtain $C_0$, $C_2$ and $C_4$ it is sufficient
to know $\gRAP(M)$ to values of $M=2$, $M=4$ and $M=16$, respectively, such that they can be determined from 
the results we computed. However, to unequivocally fix $C_7$ it is necessary to know $\gRAP(M)$ to 
values of the spin $M=128$ which is far from our reach. Nevertheless, we can assume, as a natural 
refinement of the maximal transcendentality principle, that the coefficients of the harmonic sums 
entering $C_7$ are integers. With this assumptions a result is easily found~\footnote{
As is usual in such kind of conjectures, there is a powerful numerical test that can be applied to 
any guesswork. Typically, one is able to compute spin dependent expressions like $C_7(M)$ up to a 
reasonable maximum value of $M$ in exact (rational) form. On the other hand, numerical values can be obtained 
with a very high number of digits for quite larger values of $M$. Thus, given a conjectured expression obtained 
from data up to $M_{\rm max}$ one can always test it beyond that limit with a precision of several 
hundreds of digits. These kinds of tests are always passed by the expressions we derive in this paper.}.  
The final result with all harmonic sums being of argument $M/2$ is given by
	\begin{eqnarray}\label{gwrapping}
	\gRAP(M)&=&-64 \,g^{10}\,S_{1}^{2}\big( 35\zeta_7 -40S_{2}\zeta_5 +(-8S_{4}+16S_{2,2})\zeta_3 \no\\
		&&+2S_{7}-4S_{2,5}-2S_{3,4}-4S_{4,3}-2S_{6,1}+8S_{2,2,3}+4S_{3,3,1}\big)\,.
	\end{eqnarray}
For fixed values of $M$ this result matches exactly numerical evaluations of the proposed $Y$-system~\cite{Gromov:2009tv}.

Wrapping corrections, by their nature, should not modify the leading asymptotic behavior (\ref{g10asy}). 
The result (\ref{gwrapping}) confirms this expectation, since the factor $S_{1}^{2}\sim \log^2M$ multiplies 
a linear combination of harmonic sums, which have a leading asymptotic behavior $\sim1/M^2$. The first 
wrapping contribution to the asymptotic behavior therefore only enters at order $(\log^2M/M^2)$ 
(see Appendix B, formula (\ref{gwrexp})),
\be\label{gwrasy}
\gRAP(M)\sim -\Big(768 \zeta_3-\frac{16 \pi ^4}{15} \Big)\frac{\log^2{ M}}{M^2}\,, \quad \mathrm{for} \quad M\to\infty\,.
\ee 
Thus, for large values of the spin wrapping corrections are of the same order as in the case of twist-two 
operators~\cite{bftwist2}. Further similarities with the asymptotic expansion of twist-two operators are 
discussed in Appendix B.

In the previous section we stated that the asymptotic part given in Table~\ref{fiveloops} 
\emph{is} reciprocity invariant. Hence, for the complete anomalous dimension to be reciprocity invariant, 
(\ref{gwrapping}) has to satisfy this property \emph{separately}.
Writing (\ref{gwrapping}) in terms of $\underline\Omega$ and $\Omega$ 
\be
\gRAP(M)=-64\,g^{10}\,\Omega_1^2\,\Big(35\,\zeta_7+4\,\Omega_{3,3,1}+8\,\underline{\Omega_{2,2,3}}+24\,\zeta_3\,\underline{\Omega_{2,2}}-\Omega_7\Big)\,,
\ee
one checks that this is indeed true, since according to the theorem in section 3.1 the appearing structures are all parity-invariant. 

We conclude this section giving the prediction that our conjecture given in Table \ref{fiveloops} and (\ref{gwrapping}), together with the formulas in (\ref{g2})-(\ref{g8}), give for the five-loop anomalous dimension of the simplest twist-three operator with even spin ($M=2$)
\be
 \gamma (2)= 8\,g^2-24\,g^4+136\,g^6-(920+128\,\zeta_3)\,g^8
+ 8 (833 + 144 \zeta_3 + 480 \zeta_5 - 280 \zeta_7)\,g^{10}+{\cal O}(g^{12}).
 \ee

%

\renewcommand{\theequation}{6.\arabic{equation}}
 \setcounter{equation}{0}

\section{Analytic continuation}
\label{Sec:AnCont}

As already mentioned, no direct checks of the consistency of the multi-loop anomalous 
dimension (\ref{gamma}) from its pole structure are possible.

However, it is worth to analyze the behavior of the anomalous dimension to four-loops (\ref{g2})-(\ref{g8})
and the five-loop part given in Table \ref{fiveloops} and (\ref{gwrapping}) 
at the singularity nearest to the origin, $M=-2$. 
Thus, we need the small $\omega$ expansion of general nested harmonic sums of the form 
\be
S_{a_1, \dots, a_d}(-1+\omega),\qquad a_i\in \mathbb{N}.
\ee
The analytic continuation we need can be obtained by observing that from the definition of 
harmonic sums it follows that
\be
S_{a, \mathbf{b}}(-1+\omega) = S_{a, \mathbf{b}}(\omega)-\frac{1}{\omega^a}\,S_\mathbf{b}(\omega).
\ee
This simple identity allows us to proceed by trivially expanding the {\em r.h.s} around $\omega=0$. 
This is a straightforward task, once one makes use of the general formula for the derivatives of 
nested harmonic sums~\cite{bftwist2}, and takes into account that $S_\mathbf{a}(0)=0$. 

The expansion of the $n$-th loop anomalous dimension $\gamma_n$ has the general NNLO form 
\be\label{generalNNLO}
\gamma_n = a_n\,\omega^{1-2\,n} + b_n\,\zeta_2\,\omega^{3-2\,n} + c_n\,\zeta_3\,\omega^{4-2\,n} 
	   + \cdots,\qquad a_n, b_n, c_n\in \mathbb{Q}.
\ee

Up to four loops, the explicit formulas for the two highest terms (NLO) of the analytic 
continuation are given in~\cite{KLRSV}\footnote{See Eq.~(5.12) there.}.  We recall them 
here for convenience, also adding the NNLO contribution 
\ba
\gamma_2 &=& -\frac{8}{\omega }+8 \,\zeta _2 \omega -8 \,\zeta _3 \omega ^2+\ldots, \quad
\gamma_4  = -\frac{8}{\omega ^3}+\frac{16 \,\zeta _2}{\omega }+16 \,\zeta _3+\ldots, \\
\gamma_6 &=& -\frac{8}{\omega ^5}+\frac{48 \,\zeta _2}{\omega ^3}+\frac{48 \,\zeta _3}{\omega ^2}+\ldots, \quad
\gamma_8  = -\frac{8}{\omega ^7}+\frac{80 \,\zeta _2}{\omega ^5}+\frac{80 \,\zeta _3}{\omega ^4}+\ldots\,.
\ea
In~\cite{KLRSV}, an all-loop resummation at NLO was proposed\footnote{See Eq.~(5.14) there.} and
conjectured to be valid for the asymptotic $\gABA$ part. 

At five loops, we have found for the contributions of the $\gABA$ and $\gRAP$ parts, respectively, the expressions
\be
\gABA_{10} = -\frac{136}{\omega ^9}+\frac{496 \zeta _2}{\omega ^7}-\frac{784 \zeta _3}{\omega ^6}+\ldots\,, \quad
\gRAP_{10} = \frac{128}{\omega ^9}-\frac{384 \zeta _2}{\omega ^7}-\frac{128 \zeta _3}{\omega ^6}+\ldots\,.
\ee
The analytical continuation of the complete five-loop anomalous dimension is thus given by
\be
\gamma_{10}= -\frac{8}{\omega ^9}+\frac{112 \zeta _2}{\omega ^7}-\frac{912 \zeta _3}{\omega ^6}+\cdots~.
\ee

Interestingly enough, only the above formula for the \emph{complete} anomalous dimension 
matches the proposed resummation \emph{exactly}. The latter can therefore be rewritten as
\be
\gamma\inddowns{NLO} = -8\,\frac{g^2}{\omega}\,\left(\frac{1}{1-t}-\zeta_2\,\frac{1+3\,t^2}{(1-t)^2}\,\omega^2\right), \qquad t = \frac{g^2}{\omega^2},,
\ee
with the equality valid in a perturbative sense.  It is obviously tempting to extend 
such a conjecture to NNLO, trying to resum the poles that appear in (\ref{generalNNLO}) 
with  $\zeta_3$ as a coefficient. However, the five data-points available (one for each loop) hardly 
allow for a genuine resummation. In fact, the NNLO term in the above expression is likely to 
contain enough more terms to be at least in a one-to-one correspondence with the available 
constraints. Nevertheless, we find intriguing that the following {\em simple} parameterization can be 
given, valid at five loops, 
\be
\gamma\inddowns{NNLO} = -8\,\frac{g^2}{\omega}\,\left(\frac{1}{1-t}-\zeta_2\,\frac{1+3\,t^2}{(1-t)^2}\,\omega^2
+\zeta_3\,\frac{1-5\,t+3\,t^2+t^3+128\,t^4}{(1-t)^3}\,\omega^3
\right).
\ee
\subsection*{Acknowledgments}
We benefited from discussion with Niklas Beisert, Yuri Dokshitzer, Romuald Janik, Gregory Korchemsky, Anatoly Kotikov, Giuseppe Marchesini, 
Domenico Seminara, Matthias Staudacher, Arkady A. Tseytlin and Pedro Vieira.
The work of V. Forini is supported by the Alexander von 
Humboldt Foundation. The work of T. {\L}ukowski is financed from Polish science funds during 2009-2011 as a research
project. This research was supported in part by the National Science 
Foundation, under Grant No. PHY05-51164 (S.Z.).

\appendix

\section*{Appendix A:  Harmonic sums}
\refstepcounter{section}
\def\theequation{A.\arabic{equation}}
\setcounter{equation}{0}

In this Appendix we recall some useful formulas for harmonic sums \emph{with positive indices}.
The generalization to the case of arbitrary sign for the 
indices is treated in many references, for example~\cite{Blumlein} (see also Appendix A of~\cite{bftwist2}).

The basic definition of nested harmonic sums  $S_{a_1, \dots, a_n}$ is recursive
\be
S_a(N) = \sum_{n=1}^N\frac{1}{n^{a}},~~~~~~~
S_{a, \mathbf{b}}(N) = \sum_{n=1}^N\frac{1}{n^{a}}\, S_{\mathbf b}(n),
\ee
Given a particular sum $S_\mathbf{a} = S_{a_1, \dots, a_n}$ we define
\ba
\mbox{depth}\ (S_\mathbf{a}) &=& n, \\
\mbox{transcendentality} (S_\mathbf{a}) &=& \mathbf{a}  \equiv a_1 + \cdots + a_n.
\ea
For a product of $S$ sums, we define transcendentality to be the sum of the transcendentalities of the factors.

\emph{Complementary harmonic sums} are defined recursively by 
\ba
\underline{S_a} &=& S_a, \\
\underline{S_\mathbf{a}} &=& S_\mathbf{a}-\sum_{k=1}^{\ell-1} S_{a_1,\dots, a_k}\,\underline{S_{a_{k+1},\dots, a_\ell}}(\infty),
\ea
This definition is valid when the rightmost index of $\mathbf{a}$ is not 1. Otherwise, the above recursive definition
leads to a polynomial in the formal quantity $S_1(\infty)$. In this case our {\em definition} of $\underline{S_\mathbf{a}}$ 
prescribes to set $S_1(\infty)\to 0$ in the end.
%
\section*{Appendix B:  Analysis of the asymptotic structure of $\g$ and $P$}
\refstepcounter{section}
\def\theequation{B.\arabic{equation}}
\setcounter{equation}{0}

Here we analyze the first few orders of the large $M$ expansion of the twist-three anomalous 
dimension up to five-loops and its corresponding kernel $P$~\footnote{In the case of 
higher twist $L > 2$,  anomalous dimensions occupy a band~\cite{Belitsky:2004cz}.
In this paper we have considered the \emph{minimal} anomalous dimension, see~\cite{bkp} for an asymptotic 
study of the full spectrum up to three loops.}.  

The expansions of (\ref{g2})-(\ref{g8}) to ${\cal O}(1/m^{-3})$ are given by
\ba\label{g2exp}
\g_2&=&8 \log {\bar m}+\frac{4}{m}-\frac{2}{3\, m^2}+{\cal O}\Big(\frac{1}{m^4}\Big)\,,\\\no
\g_4&=&-\frac{8}{3} \pi ^2 \log{\bar m}-8 \zeta_3 +\frac{1}{m}\Big[16 \log {\bar m}-\frac{4 \pi ^2}{3}\Big]
	   -\frac{1}{m^2}\Big[8 \log {\bar m}-\frac{2 \pi ^2}{9}-12\Big]\\\label{g4exp}
    	&&+
   	\frac{1}{m^3}\Big[\frac{8}{3} \log {\bar m}-\frac{28}{3}\Big]+{\cal O}\Big(\frac{1}{m^4}\Big)\,,\\\no
\g_6&=&\frac{88}{45} \pi ^4 \log {\bar m}-8 \zeta_5 +\frac{8}{3} \pi ^2 \zeta_3
   -\frac{1}{m}\Big[\frac{32}{3} \pi ^2 \log {\bar m}+16 \zeta_3-\frac{44 \pi ^4}{45}\Big]   \\\no
&& -\frac{1}{m^2}\Big[16\log^2{\bar m}
-\Big(32+\frac{16 \pi ^2}{3}\Big)\log{\bar m}
-8+\frac{20 \pi ^2}{3}+\frac{22 \pi ^4}{135}-8 \zeta_3
      \Big]\\ \label{g6exp}
&&+\frac{1}{m^3}\Big[16\log^2{\bar m}-\Big(64+\frac{16 \pi ^2}{9}\Big)\log{\bar m}+
   16+\frac{44 \pi ^2}{9}-\frac{8 \zeta_3}{3}
      \Big]+{\cal O}\Big(\frac{1}{m^4}\Big)\,,
\ea
\ba%
\g_8&=&-\Big(\frac{584 \pi ^6}{315}+64 \zeta_3^2\Big)\log{\bar m}-\frac{32}{15} \pi ^4 \zeta_3+\frac{8}{3} \pi ^2 \zeta_5+440 \zeta_7+\frac{1}{m}\Big[\frac{48 \pi ^4}{5}\log{\bar m}
       -\frac{292 \pi ^6}{315}+\frac{32}{3} \pi ^2 \zeta_3 \no\\
       &&-32 \zeta_3^2-16\zeta_5\Big]+\frac{1}{m^2}\Big[16 \pi ^2\log^2{\bar m}-\Big(64+32 \pi ^2+\frac{24 \pi ^4}{5}-64 \zeta_3\Big)\log{\bar m}-128-\frac{8 \pi ^2}{3}\no\\
       &&+\frac{88 \pi ^4}{15}+\frac{146 \pi ^6}{945}-32
   \zeta_3 -\frac{16}{3} \pi ^2 \zeta_3+\frac{16 \zeta_3^2}{3}+8 \zeta_5
   \Big]+\frac{1}{m^3}\Big[\frac{64}{3}\log^3{\bar m}-(96+16 \pi ^2)\log^2{\bar m}\no\\
   &&+\Big(96+\frac{176 \pi ^2}{3}+\frac{8 \pi ^4}{5}-64 \zeta_3\Big)\log{\bar m}+112-\frac{56 \pi ^2}{3}-\frac{64 \pi ^4}{15}+80 \zeta_3+\frac{16}{9}
   \pi ^2 \zeta_3-\frac{8 \zeta_5}{3}
   \Big]\no\\\label{g8exp}
&&+{\cal O}\Big(\frac{1}{m^4}\Big)\,,
\ea
where $m=\frac{M}{2}$ and ${\bar m}=m\,\exp{\EulerGamma}$. 

At five loops, the large $M$ expansion of Table~\ref{fiveloops} and \eqref{gwrapping} leads to
\ba\no
\gABA_{10}&=&  \Big(\frac{28384 \pi ^8}{14175}+\frac{128}{3} \pi ^2 \zeta_3^2 +1280 \zeta_3\zeta_5\Big)\log{\bar  m}+\frac{2048}{945} \pi ^6 \zeta_3+64 \zeta_3^3+\frac{8}{45} \pi ^4 \zeta_5-\frac{440}{3} \pi ^2 \zeta_7\\\no
%
&-& 7448\zeta_9-\frac{1}{m}\Big[\Big(\frac{9472 \pi ^6}{945}+256 \zeta_3^2\Big)\log {\bar m}-\frac{14192 \pi ^8}{14175}+\frac{448}{45} \pi ^4 \zeta_3-\frac{64}{3} \pi ^2 \zeta_3^2 -\frac{32}{3} \pi ^2 \zeta_5\\\no
&-&640 \zeta_3 \zeta_5 -880 \zeta_7 \Big]
+\frac{1}{m^2}\Big[\Big(256-\frac{272 \pi ^4}{15}-128 \zeta_3\Big)\log^2 {\bar m}
+\Big(1280+\frac{128 \pi ^2}{3}\\\no
&+&\frac{496 \pi ^4}{15}+\frac{4736 \pi ^6}{945}-128 \zeta_3-\frac{160}{3} \pi ^2 \zeta_3
+128
   \zeta_3^2-288 \zeta _5\Big)\log{\bar m}+1920+\frac{128 \pi ^2}{3}\\\no
&+&\frac{64 \pi ^4}{45}-\frac{128 \pi ^6}{21}-\frac{7096 \pi ^8}{42525}+64 \zeta_3+32 \pi ^2 \zeta_3
   +\frac{224}{45} \pi ^4 \zeta_3-208 \zeta_3^2-\frac{32}{9} \pi ^2 \zeta_3^2-32 \zeta_5\\\no
&-&\frac{16}{3} \pi ^2 \zeta_5-\frac{320}{3} \zeta_3 \zeta_5-440 \zeta_7\Big] 
   -\frac{1}{m^3}\Big[\frac{256 \pi ^2}{9} \log^3{\bar m} -\Big(128 \pi ^2\\\no
&+&\frac{272 \pi ^4}{15}-64 \zeta_3\Big)\log^2{\bar m}+\Big(768+\frac{320 \pi ^2}{3}+\frac{2752 \pi ^4}{45}+\frac{4736 \pi ^6}{2835}
   -320 \zeta_3-\frac{160}{3} \pi ^2 \zeta_3\\\no
&+&\frac{128 \zeta_3^2}{3}-288 \zeta_5\Big)\log{\bar m}
   +1536+32 \pi ^2-\frac{904 \pi ^4}{45}-\frac{1792 \pi ^6}{405}+160 \zeta_3\\\label{g10exp}
&+&\frac{208}{3} \pi ^2 \zeta_3+\frac{224}{135}
   \pi ^4 \zeta_3-\frac{496 \zeta_3^2}{3}+96 \zeta_5-\frac{16}{9} \pi ^2 \zeta_5-\frac{440 \zeta_7}{3}
   \Big]+{\cal O}\Big(\frac{1}{m^4}\Big)\,,
\\\label{gwrexp}
\gRAP&=&-\Big(768 \zeta_3-\frac{16 \pi ^4}{15} \Big)\frac{\log^2{\bar m}}{m^2}+\Big(768 \zeta_3-\frac{16 \pi ^4}{15} \Big)(\log^2{\bar m}-\log{\bar m})\frac{1}{m^3}+{\cal O}\Big(\frac{1}{m^4}\Big)
\ea

As expected in the case of minimal anomalous dimension for operators of twist $L\leq 3$, logarithmic 
enhancements in the asymptotic expansions of $\gamma$ are always positive in 
power~\footnote{Terms with negative powers of the logarithm appear for twist $L\leq 3$, but for \emph{non-minimal} 
anomalous dimensions~\cite{ggg}. For twist $L>3$ terms $\sim \frac{1}{\log^p M}$ are present both for 
large $L$ and $M$~\cite{bgk} and for finite twist, see a similar discussion in~\cite{BFTT} and 
reference therein. For a general method to derive higher order terms in  the $1/M$  expansion 
at fixed $L$  see  \cite{bc}.}. 

Notice that, when expressed in terms of the variable $M=2m$, the maximal logarithmic 
terms $\log^pm/m^p$  in the expansions up to four loops, formulas (\ref{g2exp})-(\ref{g8exp}), 
are compatible with a resummation of type 
\be\label{leadinglogs}
\g(M)=f(g)\log\Big(M+\textstyle{\frac{1}{2}}f (g)\log M+\dots\Big)+\dots~.
\ee
According to this, their coefficients are simply proportional to $f^{m+1}$
\be\label{leadinglogs2}
\g(M)\sim f\, \log M+\frac{f^2}{2}\,\frac{\log M}{M}-\frac{f ^3}{8}\,\frac{\log^2 M}{M^2}+...
\ee
where $f$ is the universal scaling function, whose weak coupling expansion to five-loop 
order can be found in~\cite{bes}.
At five loops, the pattern (\ref{leadinglogs2}) is broken by the term $\log^2M/M^2$ in the 
expansion (\ref{g10exp}) above~\footnote{Further maximal logarithmic terms as 
$\log^3M/M^3,\log^4M/M^4$ continue to obey the rule  as dictated by further orders 
in (\ref{leadinglogs2}). Coefficients of  terms $\log^pM/M^p$ with $p>4$ are absent in the expansion, as checked up to $1/M^{25}$. This is again consistent with (\ref{leadinglogs2}),  being such coefficients of the form $\frac{f^6}{160},-\frac{f^7}{384},\dots$, they would  contribute starting at 6 loops.}.  Interestingly, it is precisely at this order in the large $M$ 
expansion that wrapping corrections start to contribute. Explicitly, while on the basis 
of (\ref{leadinglogs2}) one would expect at five loops a term of type
\be\label{predc22}
(c_{22})_5\frac{\log^2M}{M^2}~~~~~~~~{\rm with}~~~~~~~~(c_{22})_5^{\rm naive}=\big(\textstyle{-\frac{f^3}{8}}\big)_5=-\frac{1024}{15}\,\pi^4=-6144\, \zeta_4~,
\ee
reexpressing (\ref{g10exp})  and (\ref{gwrexp})  in terms of $M$ one finds 
\be\label{effc22}
(c_{22})_5^{\rm ABA}=1024 -512 \zeta_3- 6528 \zeta_4~~~~~~~~{\rm and}~~~~~~~~
(c_{22})_5^{\rm wrapping}=- 3072\,\zeta_3+384\,\zeta_4\,.
\ee
The sum of these terms does clearly not reproduce (\ref{predc22}). This is analogous 
to the case of twist two operators~\cite{bftwist2}~\footnote{One difference is however 
that the degree of transcendentality of the asymptotic and wrapping contributions, that differs 
in (\ref{effc22}), is the same in the twist-two case.}.
 
It is interesting to notice that the structure, lost at higher orders in $1/m$, of the first 
terms in the expansion for $\gRAP$ is also present in its twist-two analogue. The appearance of an 
overall coefficient multiplying the $1/m^2$ and $1/m^3$ terms was already noticed in formula 
(C.5) of~\cite{bftwist2}. Another analogy  between the leading  asymptotic behavior of twist-2 and 
twist-3 wrapping contributions is their negative sign and their common pattern 
$\sim c_n \,\zeta_n+c_{n+1}\,\zeta_{n+1}$, where $n$ coincides with the twist.

Concerning the $P$-kernel, the logarithmic structure 
to four loops is remarkably  \emph{simpler} than the one of the corresponding anomalous 
dimension, as it is only \emph{linear} in $\log M$~\cite{bdm}. In particular, there are 
no maximally enhanced terms of the form $(\log M/M)^k$. As discussed in~\cite{bk,dok2}, this 
feature of $P$ translates into the chance of a resummation of type (\ref{leadinglogs}). 
This asymptotic structure changes at five loops.

The $P$ function, derived by inverting formula (\ref{nonlinear}), reads  
in terms of $m=\frac{M}{2}$ ($\partial\equiv\partial_m$) to five-loops
\be
{\cal P}(m) = \sum_{k=1}^\infty \frac{1}{k!}\left(\textstyle{-\frac{1}{4}}\partial\right)^{k-1}[\gamma(m)]^k =  \gamma-\frac{1}{8}\,(\gamma^2)'+\frac{1}{96}\,(\gamma^3)''-\frac{1}{1536}\,(\gamma^4)''' +\frac{1}{30720}\,(\g^5)''''+\cdots .\nonumber
\ee
Replacing $\g$ by the perturbative expansion (\ref{gamma}) we can formally write at five loops 
\ba\label{P10formal}
P_{10}=\g_{10} - 
 \frac{1}{4} ( \gamma_{4}  \gamma_{6} + \gamma_{2} \gamma_{8} )' + 
 \frac{1}{32} ( \gamma_{2} \gamma_{4}^2 + \gamma_{2}^2  \gamma_6 )'' 
 - \frac{1}{384} ( \gamma_2^3 \gamma_4 )''' +  \frac{1}{30720} (\gamma_2^5 )'''' \,.
\ea
Expanded at large $M$, including the wrapping contribution, this becomes
\ba\no
P_{10}&=&\Big(\frac{28384 \pi ^8}{14175}+\frac{128}{3} \pi ^2 \zeta_3^2 +1280 \zeta_3\zeta_5\Big)\log{\bar  m}
+\frac{2048}{945} \pi ^6 \zeta_3+64 \zeta_3^3+\frac{8}{45} \pi ^4 \zeta_5-\frac{440}{3} \pi ^2 \zeta_7\\\no
%
&-& 7448\zeta_9
+\frac{1}{m}\Big[\frac{14192 \pi ^8}{14175}+\frac{64}{3} \pi ^2 \zeta_3^2 +640 \zeta_3
   \zeta_5\Big]+\frac{1}{m^2}\Big[\Big(256-896 \zeta_3\Big)\log^2{\bar m}+\\\no
&+& \Big(1280+\frac{128 \pi ^2}{3}-\frac{16 \pi ^4}{15}-128 \zeta_3-\frac{64}{3}
   \pi ^2 \zeta_3 -320 \zeta_5\Big)\log{\bar m}+1920+\frac{128 \pi ^2}{3}+\frac{64 \pi ^4}{45}+
   \\\no
&-&\frac{1024 \pi
   ^6}{945}-\frac{7096 \pi ^8}{42525}+64 \zeta_3-64 \zeta_3^2-\frac{32}{9} \pi ^2 \zeta_3^2-\frac{320}{3} \zeta_3 \zeta_5\Big]\\\no
&-&\frac{1}{m^3}\Big[(256-896 \zeta_3)\log^2{\bar m}+\Big(1024+\frac{128 \pi ^2}{3}-\frac{16 \pi ^4}{15}+768 \zeta_3-\frac{64}{3} \pi ^2 \zeta_3-320 \zeta_5\Big)\log{\bar m}\\\label{P10exp}
&+&1280+\frac{64 \pi ^2}{3}+\frac{88 \pi ^4}{45}-\frac{1024 \pi ^6}{945}+128
   \zeta_3+\frac{32}{3} \pi ^2 \zeta_3-64 \zeta_3^2+160 \zeta_5\Big]+{\cal O}\Big(\frac{1}{m^4}\Big)\,.
\ea
The ``simplicity''  feature is lost, because at order $1/m^2$  a term $\log^2 m/m^2$ appears, 
which is responsible for the above formula (\ref{effc22}).

We recall that the consequences (\ref{leadinglogs}) and (\ref{leadinglogs2}) of the simplicity 
of the $P$ function \emph{and} the knowledge of $f$ to presumably all loops~\cite{bes} allow in 
principle an all-loop prediction for such maximal logarithmic terms, whose coefficients should 
be simply proportional to $f^{m+1}$. Indeed, such \emph{inheritance} has been checked at strong 
coupling in~\cite{BFTT} up to one-loop in the semiclassical sigma model expansion, as well as 
in~\cite{Ishizeki:2008tx} at the classical level. An independent strong coupling confirmation 
of (\ref{leadinglogs}) for twist-two operators has recently been given in~\cite{Freyhult:2009my}. To clarify if and how the difference in the simplicity of the $P$ at weak and strong coupling works, further orders  in the semiclassical sigma model expansion would be needed.

We conclude the appendix by reporting the separate contributions to $P_{10}$ coming from the dressing factor, which
obey the property described in point 4 of Section~4. They read
\ba\no
P_{10}^{(\zeta_3)}&\equiv&\g_{10}^{(\zeta_3)}-\frac{1}{4}(\g_2\,\g_8^{(\zeta_3)})'=896 S_ {6} - 2304 S_ {1, 5} - 1792 S_ {2, 4} - 768 S_ {3, 3} - 
 1792 S_ {4, 2} - 2304 S_ {5, 1} \\\no
 &&+ 2560
   S_ {1, 1, 4} + 512 S_ {1, 2, 3} + 1536 S_ {1, 3, 2} + 
 3584 S_ {1, 4, 1} + 512 S_ {2, 1, 3} + 1536
   S_ {2, 3, 1}+ 512 S_ {3, 1, 2} \\\no
   && + 512 S_ {3, 2, 1} + 
 2560 S_ {4, 1, 1} - 2048 S_ {1, 1, 3, 1} - 2048S_ {1, 3, 1, 1}+512\zeta_2 S_ {1} S_ {3} - 512 S_ {1} S_ {2} S_ {3} \\
   && + 
 768 \zeta_4 S_ {1}^2 - 768 S_ {1}^2 S_ {4}\,,
 \\
P_{10}^{(\zeta_5)}&\equiv&\g_{10}^{(\zeta_5)}=1280\,S_1\,S_3\,.
\ea
In the first line we included the dressing-induced contribution at four loops, which is proportional to $\zeta_3$. 
These contributions can be expressed in terms of parity invariant combinations (see Theorem (c) in Section 3.1) as
\ba\label{P3RR}
P_{10}^{(\zeta_3)}&=&256(8\,\Omega_1\,\Omega_{1,1,3}-4\Omega_1^2\,\Omega_{1,3}+\Omega_{3,3}+S_1\,\Omega_5 +2\zeta_2 \Omega_1\, \Omega_3+3\zeta_4\,\Omega_1^2)\,,\\\label{P5RR}
P_{10}^{(\zeta_5)}&=&1280\,\Omega_1\,\Omega_3.
\ea



\begin{thebibliography}{99}
\bibitem{Integrable Structures}
 J.~A.~Minahan and K.~Zarembo,
 JHEP {\bf 0303} (2003) 013
 [arXiv:hep-th/0212208].
 $\bullet$
 I.~Bena, J.~Polchinski and R.~Roiban,
 Phys.\ Rev.\  D {\bf 69}, 046002 (2004),
 [arXiv:hep-th/0305116].
 $\bullet$
  N.~Beisert, C.~Kristjansen and M.~Staudacher,
  Nucl.\ Phys.\  B {\bf 664}, 131 (2003)
  [arXiv:hep-th/0303060].

\bibitem{Beisert:2005fw}
  N.~Beisert and M.~Staudacher,
  Nucl.\ Phys.\  B {\bf 727}, 1 (2005),
  [arXiv:hep-th/0504190].

\bibitem{factorized}
  M.~Staudacher,
  JHEP {\bf 0505}, 054 (2005),
  [arXiv:hep-th/0412188].

\bibitem{Beisert:2005tm}
  N.~Beisert,
  Adv.\ Theor.\ Math.\ Phys.\  {\bf 12}, 945 (2008),
  [arXiv:hep-th/0511082].

\bibitem{Janik:2006dc}
 R.~A.~Janik,
 Phys.\ Rev.\ D {\bf 73} (2006) 086006,
 [arXiv:hep-th/0603038].

\bibitem{Beisert:2006ib}
  N.~Beisert, R.~Hernandez and E.~Lopez,
  JHEP {\bf 0611} (2006) 070,
  [arXiv:hep-th/0609044].

 \bi{bes}
N.~Beisert, B.~Eden and M.~Staudacher,
  J.\ Stat.\ Mech.\  {\bf 0701}, P021 (2007),
  [arXiv:hep-th/0610251].

\bibitem{Bern:2006ew}
  Z.~Bern, M.~Czakon, L.~J.~Dixon, D.~A.~Kosower and V.~A.~Smirnov,
  Phys.\ Rev.\  D {\bf 75}, 085010 (2007),
  [arXiv:hep-th/0610248].
$\bullet$
  F.~Cachazo, M.~Spradlin and A.~Volovich,
  Phys.\ Rev.\  D {\bf 75}, 105011 (2007),
  [arXiv:hep-th/0612309].

\bibitem{Rej:2007vm}
  A.~Rej, M.~Staudacher and S.~Zieme,
  J.\ Stat.\ Mech.\  {\bf 0708}, P08006 (2007),
  [arXiv:hep-th/0702151].
$\bullet$
  K.~Sakai and Y.~Satoh,
  Phys.\ Lett.\  B {\bf 661}, 216 (2008),
  [arXiv:hep-th/0703177].
$\bullet$
  R.~A.~Janik and T.~Lukowski,
  Phys.\ Rev.\  D {\bf 78}, 066018 (2008),
  [arXiv:0804.4295 [hep-th]].

\bibitem{Ambjorn:2005wa}
  J.~Ambjorn, R.~A.~Janik and C.~Kristjansen,
  Nucl.\ Phys.\  B {\bf 736}, 288 (2006),
  [arXiv:hep-th/0510171].

\bibitem{KLRSV}
  A.~V.~Kotikov, L.~N.~Lipatov, A.~Rej, M.~Staudacher and V.~N.~Velizhanin,
  J.\ Stat.\ Mech.\  {\bf 0710}, P10003 (2007),
  [arXiv:0704.3586 [hep-th]].

\bibitem{Fiamberti:2007rj}
  F.~Fiamberti, A.~Santambrogio, C.~Sieg and D.~Zanon,
  Phys.\ Lett.\  B {\bf 666}, 100 (2008).
  [arXiv:0712.3522 [hep-th]].

\bibitem{Bajnok:2008bm}
  Z.~Bajnok and R.~A.~Janik,
  Nucl.\ Phys.\  B {\bf 807}, 625 (2009),
  [arXiv:807.0399 [hep-th]].

\bibitem{Velizhanin:2008jd}
  V.~N.~Velizhanin,
  [arXiv:0808.3832 [hep-th]].

\bibitem{Luscher:1985dn}
  M.~Luscher,
  Commun.\ Math.\ Phys.\  {\bf 104} (1986) 177.

\bibitem{Janikprev}
  R.~A.~Janik and T.~Lukowski,
  Phys.\ Rev.\  D {\bf 76}, 126008 (2007),
  [arXiv:0708.2208 [hep-th]].

\bibitem{Bajnok:2008qj}
  Z.~Bajnok, R.~A.~Janik and T.~Lukowski,
  [arXiv:0811.4448 [hep-th]].

\bibitem{bftwist2}
  M.~Beccaria and V.~Forini,
  [arXiv:0901.1256 [hep-th]].

\bibitem{Veliz}
  V.~N.~Velizhanin,
  [arXiv:0811.0607 [hep-th]].

\bibitem{Arutyunov:2007tc}
  G.~Arutyunov and S.~Frolov,
  JHEP {\bf 0712}, 024 (2007),
  [arXiv:0710.1568 [hep-th]].
$\bullet$
  G.~Arutyunov and S.~Frolov,
  [arXiv:0901.1417 [hep-th]].

\bibitem{Gromov:2008gj}
  N.~Gromov, V.~Kazakov and P.~Vieira,
  [arXiv:0812.5091 [hep-th]].

\bibitem{Gromov:2009tv}
  N.~Gromov, V.~Kazakov and P.~Vieira,
  [arXiv:0901.3753 [hep-th]].

\bibitem{Kotikov:2002ab}
  A.~V.~Kotikov and L.~N.~Lipatov,
  Nucl.\ Phys.\  B {\bf 661}, 19 (2003),
  [Erratum-ibid.\  B {\bf 685}, 405 (2004)],
  [arXiv:hep-ph/0208220].

\bibitem{Freyhult:2009my}
  L.~Freyhult and S.~Zieme,
  [arXiv:0901.2749 [hep-th]].

\bibitem{Fioravanti:2009xt}
  D.~Fioravanti, P.~Grinza and M.~Rossi,
  [arXiv:0901.3161 [hep-th]].

\bibitem{BDS}
 N.~Beisert, V.~Dippel and M.~Staudacher,
 JHEP {\bf 0407} (2004) 075,
 [arXiv:hep-th/0405001].

\bibitem{Kotikov:2008pv}
  A.~V.~Kotikov, A.~Rej and S.~Zieme,
  [arXiv:0810.0691 [hep-th]].

\bibitem{Belitsky:2006wg}
  A.~V.~Belitsky, G.~P.~Korchemsky and D.~Mueller,
  Nucl.\ Phys.\  B {\bf 768} (2007) 116,
  [arXiv:hep-th/0605291].
  $\bullet$
  A.~V.~Belitsky,
  Phys.\ Lett.\  B {\bf 643} (2006) 354,
  [arXiv:hep-th/0609068].

\bibitem{Beccaria:2007cn}
  M.~Beccaria,
  JHEP {\bf 0706}, 044 (2007),
  [arXiv:0704.3570 [hep-th]].
  
\bibitem{Beccaria:2007gu}
 M.~Beccaria and V.~Forini,
  ``Anomalous dimensions of finite size field strength operators in N=4 SYM,''
  JHEP {\bf 0711}, 031 (2007)
  [arXiv:0710.0217 [hep-th]].

\bi{dok1}
  Yu.~L.~Dokshitzer, G.~Marchesini and G.~P.~Salam,
  Phys.\ Lett.\  B {\bf 634}, 504 (2006),
  [arXiv:hep-ph/0511302].

 \bi{dok2}
  Yu.~L.~Dokshitzer and G.~Marchesini,
  Phys.\ Lett.\  B {\bf 646}, 189 (2007),
  [arXiv:hep-th/0612248].

\bibitem{bk}
  B.~Basso and G.~P.~Korchemsky,
  Nucl.\ Phys.\  B {\bf 775}, 1 (2007),
  [arXiv:hep-th/0612247].

\bibitem{bf}
  M.~Beccaria and V.~Forini,
 JHEP {\bf 0806}, 077(2008),
  [arXiv:0803.3768[hep-th]]
  
\bibitem{Forini:2008ky}
  V.~Forini and M.~Beccaria,
   Nonlinear Physics. Theory and Experiment. V, Gallipoli (Italy), June 12-21, 2008,
   [arXiv:0810.0101 [hep-th]].

\bibitem{bdm}
  M.~Beccaria, Yu.~L.~Dokshitzer and G.~Marchesini,
  Phys.\ Lett.\  B {\bf 652}, 194 (2007),
  [arXiv:0705.2639 [hep-th]].

\bibitem{Beccaria:2007pb}
  M.~Beccaria,
  JHEP {\bf 0709}, 023 (2007),
  [arXiv:0707.1574 [hep-th]].
  
\bibitem{Belitsky:2003ys}
  A.~V.~Belitsky, A.~S.~Gorsky and G.~P.~Korchemsky,
  Nucl.\ Phys.\  B {\bf 667}, 3 (2003)
  [arXiv:hep-th/0304028].

\bibitem{Eden:2006rx}
  B.~Eden and M.~Staudacher,
  J.\ Stat.\ Mech.\  {\bf 0611}, P014 (2006),
  [arXiv:hep-th/0603157].
  




\bibitem{Freyhult:2007pz}
  L.~Freyhult, A.~Rej and M.~Staudacher,
  J.\ Stat.\ Mech.\  {\bf 0807}, P07015 (2008),
  [arXiv:0712.2743 [hep-th]].

\bibitem{Bombardelli:2008ah}
  D.~Bombardelli, D.~Fioravanti and M.~Rossi,
  [arXiv:0802.0027 [hep-th]].

\bibitem{Blumlein}
  J.~Blumlein,
  Comput.\ Phys.\ Commun.\  {\bf 159}, 19 (2004)
  [arXiv:hep-ph/0311046].

\bibitem{Belitsky:2004cz}
  A.~V.~Belitsky, V.~M.~Braun, A.~S.~Gorsky and G.~P.~Korchemsky,
  Int.\ J.\ Mod.\ Phys.\  A {\bf 19}, 4715 (2004)
  [arXiv:hep-th/0407232].

  \bibitem{bkp}
  A.~V.~Belitsky, G.~P.~Korchemsky and R.~S.~Pasechnik,
  [arXiv:0806.3657 [hep-ph]].

\bibitem{ggg}
V.~M.~Braun, S.~E.~Derkachov, G.~P.~Korchemsky and A.~N.~Manashov,
  Nucl.\ Phys.\  B {\bf 553}, 355 (1999)
  [arXiv:hep-ph/9902375]. 
$\bullet$
S.~E.~Derkachov, G.~P.~Korchemsky and A.~N.~Manashov,
  Nucl.\ Phys.\  B {\bf 566}, 203 (2000)
  [arXiv:hep-ph/9909539].

 \bi{bgk}
  A.~V.~Belitsky, A.~S.~Gorsky and G.~P.~Korchemsky,
  { Logarithmic scaling in gauge / string correspondence},
  Nucl.\ Phys.\  B {\bf 748}, 24 (2006)
  [arXiv:hep-th/0601112].

\bibitem{BFTT}
  M.~Beccaria, V.~Forini, A.~Tirziu and A.~A.~Tseytlin,
 Nucl.\ Phys.\ B, In Press, arXiv:0809.5234 [hep-th].

  \bibitem{bc}
  M.~Beccaria and F.~Catino,
  JHEP {\bf 0801}, 067 (2008)
  [arXiv:0710.1991 [hep-th]].

\bibitem{Ishizeki:2008tx}
  R.~Ishizeki, M.~Kruczenski, A.~Tirziu and A.~A.~Tseytlin,
  [arXiv:0812.2431 [hep-th]].


\end{thebibliography}
\end{document}